\documentclass[12pt]{article}
\usepackage{amsmath,latexsym}
\usepackage{graphicx}
\begin{document}

 \title{\Huge Wormhole geometry from real feasible matter sources
  }
 \author{F.Rahaman$^*$, M.Kalam$^{\dag}$
 and K A Rahman$^*$}

\date{}
 \maketitle
 \begin{abstract}
We provide a  prescription of real feasible sources that  supply
fuel to construct a traversable wormhole. A class of exact
solutions for Einstein-Maxwell field equations describing wormhole
with an anisotropic matter distribution has  been presented. The
anisotropy plays a crucial role for the existence of the spacetime
comprising wormhole geometry.

\end{abstract}

  \footnotetext{
 $*$Dept.of Mathematics, Jadavpur University, Kolkata-700 032, India

                                  E-Mail:farook\_rahaman@yahoo.com\\
$\dag$Dept. of Phys. , Netaji Nagar College for Women, Regent Estate, Kolkata-700092, India.\\
E-Mail:mehedikalam@yahoo.co.in\\
}
    \mbox{} \hspace{.2in}

\title{ \underline{\textbf{Introduction}}: }

During last 30 years, more than 700 articles [1] have been written
on wormholes after the pioneering work of Morris and Thorne [2].
Scientists show interest on wormhole physics because it  opens up
a possibility to construct a time machine [3]. It  is well known
that wormhole demands exotic matter ( matter that violates  weak
or null energy conditions [WEC or NEC] ). There are several
proposals have been proposed  regarding the exotic matter sources
 that support wormhole spacetime. But scientists remain silent
 whether it is possible to manufacture or creation of this exotic
 matter. As exotic matter violates the NEC, so it is quite
 impossible to provide such type of matter source to construct a
 wormhole.  In other words, time machine ( which is a possible
 consequence of wormhole ) can never be constructed. In this
 letter we will give a  clue how  one can get feasible matter
 sources  that supply fuel to construct and sustain a  wormhole.
In the  present investigation, we will show that if we are
supplied anisotropic matter source and electromagnetic field, then
one could construct a traversable wormhole.

\pagebreak

 Anisotropic matter sources are not absurd ( rather we
say, not exotic ). Recent observations on highly compact
astrophysical objects like X ray pulsar Her X-1, X ray buster 4U
1820-30, millisecond pulsar SAX J 1808.4 - 3658,   X ray sources
4U 1728 - 34 etc indicate that the densities of such objects are
normally beyond  nuclear matter density. Theoretical advances in
the last decades indicate that pressures within such stars are
anisotropic i.e. radial pressure ($p_r$) is not equal to
tangential pressure ($p_t$) in such bodies [4].

If an advanced engineer imbued with new ideas will able to produce
anisotropic matter source ( like above astrophysical compact
objects ), then  wormhole could be constructed physically. Rather
producing negative energy or matter that violates NEC, we think,
it is simpler to creation    compact object with  density greater
than the nuclear matter density.

\title{ \underline{\textbf{Basic equations for constructing wormholes}}: }

For the  present study the metric for static spherically symmetric
spacetime is taken as
\begin{equation}
               ds^2=  - e^{\nu(r)} dt^2+ e^{\mu(r)} dr^2+r^2( d\theta^2+sin^2\theta
               d\phi^2)
         \label{Eq3}
          \end{equation}
          The most general energy momentum tensor compatible with
          spherically  symmetry is
\begin{equation}
               T_\nu^\mu=  ( \rho + p_t)u^{\mu}u_{\nu} - p_t g^{\mu}_{\nu}+ (p_r -p_t )\eta^{\mu}\eta_{\nu}
         \label{Eq3}
          \end{equation}
with

 $u^{\mu}u_{\mu} = - \eta^{\mu}\eta_{\mu} = 1 $.

          The Einstein-Maxwell field equations for the above spherically
          symmetric metric corresponding to the charged
          anisotropic matter distribution  are given by
\begin{equation}e^{-\mu}
[\frac{\mu^\prime}{r} - \frac{1}{r^2} ]+\frac{1}{r^2}= 8\pi \rho +
E^2
\end{equation}
\begin{equation}e^{-\mu}
[\frac{1}{r^2}+\frac{\nu^\prime}{r}]-\frac{1}{r^2}= 8\pi p_r - E^2
\end{equation}
\begin{equation}e^{-\mu}
[\frac{1}{2}(\nu^\prime)^2+ \nu^{\prime\prime}
-\frac{1}{2}\mu^\prime\nu^\prime + \frac{1}{r}({\nu^\prime-
\mu^\prime})] =8\pi p_t + E^2 \end{equation} and
\begin{equation}(r^2E)^\prime = 4\pi r^2 \sigma e^{\frac{\mu}{2}}
\end{equation}
Equation (6) can equivalently be expressed in the form
\begin{equation} E(r) = \frac{1}{r^2}\int_0^r 4\pi r^2 \sigma
e^{\frac{\mu}{2}}dr = \frac{q(r)}{r^2}
\end{equation}
where $q(r)$ is the total charge of the sphere under
consideration.

\pagebreak

Also, the conservation equation is given by

\begin{equation} \frac{dp_r}{dr}
+ ( \rho + p_r ) \frac{ \nu^\prime }{2} = \frac{1}{ 8 \pi
r^4}\frac{dq^2}{dr} + \frac{2(p_t - p_r)}{r}
\end{equation}
Here, $ \rho, E, \sigma$   and  $ q $ are respectively the matter
energy density, electric field strength, electric  charged density
and electric charge and $ \Delta = p_t - p_r  $, is the measure of
anisotropy. The prime denotes derivative with respect to 'r'.

\title{ \underline{\textbf{Solutions}}: }
To control   the solutions, we assume the following assumptions:

(a) \begin{equation} \nu(r) = 0 \end{equation}

\textbf{Argument:} One of the traversability properties is that
the tidal gravitational forces experienced by a traveller must be
reasonably small. So, we assume a zero tidal force as seen by the
stationary observer. Thus one of the traversability conditions is
automatically satisfied.

(b) \begin{equation}  \Delta = p_t - p_r  = -\alpha q^2 r^n
\end{equation}
where, $\alpha > 0 $ and n are arbitrary constants.

 \textbf{Argument:} According to Usov [5], the stars whose density
 beyond the nuclear matter density, the anisotropy within such
 stars could be the presence of strong electromagnetic field. So,
 the above assumption is justified.

 (c) \begin{equation}  \sigma e^{\frac{\mu}{2}} = \sigma_0 r^s
\end{equation}
( $\sigma_0$  and  s  are  arbitrary constants  )

 \textbf{Argument:}
In usual sense, the term $  \sigma e^{\frac{\mu}{2}} $ occurring
inside the integral sign in the equation (7), is called the volume
charge density and hence the condition $  \sigma e^{\frac{\mu}{2}}
= \sigma_0 r^s $ , can equivalently be interpreted as the  volume
charge density being polynomial function of 'r'. The constant $
\sigma_0  $ is the charge density at $ r = 0 $, the center of the
charged matter [6].

(d) \begin{equation}  p_r = m\rho
\end{equation}

 \textbf{Argument:}
The above equation indicates the equation  of state with $0 < m  <
1 $.

\pagebreak

 Taking into account of equations (9) - (12), one  gets
the following solutions of the field equations (3)  - (8) as

\begin{equation}  q^2(r)  =  \frac{16\pi^2\sigma_0^2}{(s+3)^2}
r^{2s+ 6}
\end{equation}

\begin{equation}  E^2(r)  =  \frac{16\pi^2\sigma_0^2}{(s+3)^2}
r^{2s+ 2}
\end{equation}

\begin{equation}  \rho  =  A r^{2s+2} + B r^{2s+n+6}
\end{equation}
where  $ A = \frac{4\pi\sigma_0^2}{m (s +  3 )(2s +  2 )}$ and   $
B = \frac{-32\pi^2  \alpha  \sigma_0^2}{m (s +  3 )^2(2s +  n+6
)}$ .

\begin{equation}  e^{-\mu}   =  1 - \frac{b(r)}{r}
\end{equation}
where,
\begin{equation}  b(r)  =  F r^{u+1} + X
r^{w+1}
\end{equation}
where, $  w = 2s+4 $, $ u= 2s+n+8$ , $ X = \frac{32\pi^2
\sigma_0^2} {m( s+3 )( 2s+2 )( 2s+5 )} +
 \frac{16\pi^2 \sigma_0^2} {( s+3 )^2( 2s+5 )}$
and \linebreak
$ F = \frac{  - 256\pi^3 \alpha  \sigma_0^2}{m( s+3
)^2(2s + n+6 )(2s + n+9 )}$

\title{ \underline{\textbf{Wormhole structure}}: }

For the assumption $\nu(r) = 0 $ implies no horizon exists in the
spacetime.
 Since the space time is asymptotically flat
i.e. $\frac{b(r)}{r}\rightarrow 0 $ as $ \mid r \mid \rightarrow
\infty $, the Eq.{(17)} is consistent only when $2s +n + 8 < 0  $
and $ 2s + 4 < 0  $.

Also, one can note that X is negative for the above restrictions
and  as $ \mid r \mid \rightarrow \infty $, $q^2(r)$ and $E^2(r)$
$ \rightarrow 0 $, so one has to take the following restrictions
on 's' to get wormhole structure as $   s < -3, 2s +n + 8 < 0$ but
$ 2s +n + 9 > 0 $.

Here the throat occurs at $ r= r_0 $ for which $ b(r_0) = r_0 $
i.e. $ 1 = F r_0^u + X r_0^w$. For the suitable choices of the
parameters, the graph of the function $ G(r) = b (r) - r$
indicates the point $r_0$ , where G(r) cuts the 'r' axis (see Fig.
1 ).  From the graph, one can also note that  when $r>r_0 $,
$G(r)< 0$ i.e. $  b(r) -r < 0 $. This implies $ \frac{b(r) }{r} <
1 $ when $r>r_0 $. Also,  from the graph, we see that G is a
decreasing function of r and hence $G^{\prime}(r)< 0$. In other
words, $b^{\prime}(r_0)< 1$  i.e. flare-out condition has been
satisfied.
\begin{figure}[htbp]
    \centering
        \includegraphics[scale=.8]{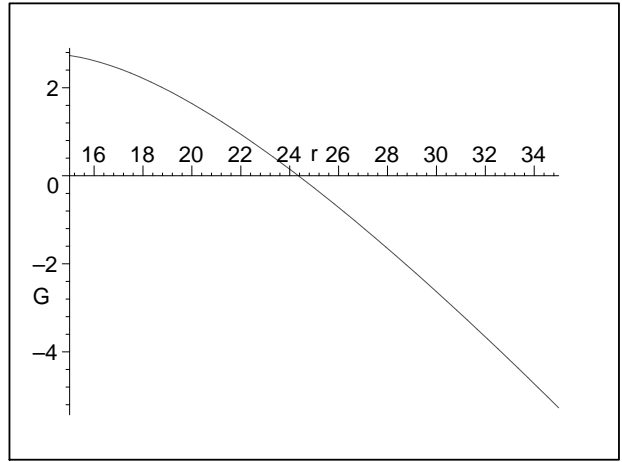}
        \caption{ Throat occurs where $G(r)$ cuts 'r' axis
        ( choosing suitably the parameters as $\sigma_0 = 1,
    \alpha = .0001 , s = -3.5, n = -1.5, m = .5 $ ).}
    \label{fig:1}
\end{figure}

\pagebreak

 Thus our solution describing a static spherically
symmetric wormhole supported by anisotropic matter distribution in
presence of electromagnetic field. The axially symmetric embedded
surface $ z = z(r)$ shaping the Wormhole's spatial geometry is a
solution of
\begin{equation}\label{Eq21}
 \frac{dz}{dr}=\pm \frac{1}{\sqrt{\displaystyle{\frac{r}{b(r)}}-1}}
 \end{equation}
  One can note from the definition of Wormhole that at   $ r= r_0 $
  (the wormhole throat) Eq.(18) is divergent i.e.  embedded surface is
   vertical there.
One can see that embedding diagram of this wormhole ( expand
binomially in powers of r and retaining a few terms)
 in Fig-2.
 The surface of revolution of the curve about the vertical z axis makes
 the diagram complete (see  Fig.3).
\begin{figure}[htbp]
    \centering
        \includegraphics[scale=.8]{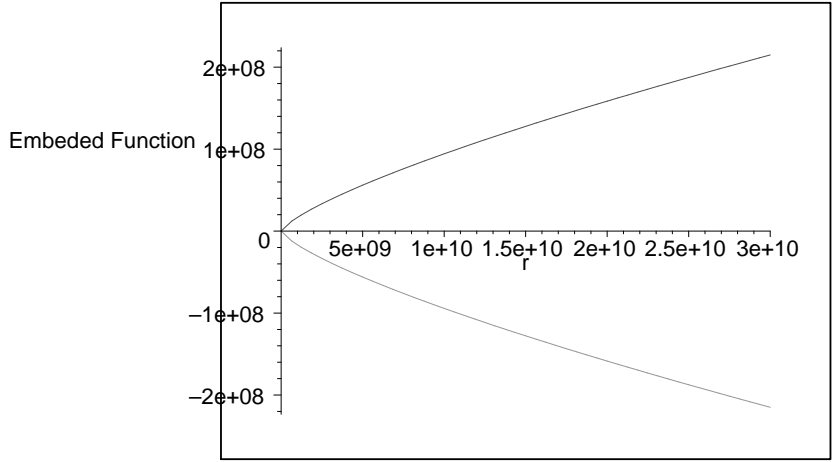}
    \caption{The embedding diagram of the wormhole( retaining a
few terms and choosing  suitably the
        parameters as  $\sigma_0 = 1,
    \alpha = .0001 , s = -3.5, n = -1.5, m = .5 $ ).   }
    \label{fig:wormhole}
\end{figure}

\pagebreak

\begin{figure}[htbp]
    \centering
        \includegraphics[scale=.8]{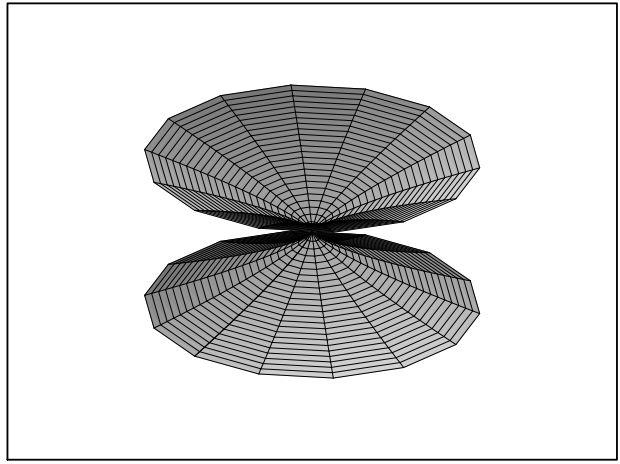}
    \caption{The full visualization of the surface generated by the rotation of the embedded
    curve (  retaining a
few terms ) about the vertical z axis.  }
    \label{fig:wormhole}
\end{figure}

\title{ \underline{\textbf{Final Remarks}}: }

In  this article, we provide matter sources that give birth to
wormhole like geometry. By considering the anisotropy of  the
matter distribution, we have shown how wormhole could be
constructed in presence of electromagnetic field. We have
presented  an example
 of the possible structure of the wormhole  generated by the rotation of the embedded
    curve  about the vertical z axis.
 One can see that $ \rho_{effective} > 0, $ $ \rho_{effective}$ +
$p_{r}$ $_{effective}> 0 $, $ \rho_{effective}$ + $p_{t}$
$_{effective}> 0 $ for all  $ r > r_0 $ i.e. all energy conditions
are satisfied out side the wormhole throat. But at the throat i.e.
at $r = r_0$, NEC is violated. Nevertheless, this wormhole has
been constructed by real feasible matter sources. Although several
assumptions were considered, we give valid arguments against it.
We hope scientists would be motivated by our approach and in
future, will try to find sophisticated  way for constructing
feasible wormhole.

\pagebreak

 { \bf Acknowledgments }

          F.R is thankful to Jadavpur University and DST , Government of India for providing
          financial support. MK has been partially supported by
          UGC,
          Government of India under MRP scheme.  \\


\begin{thebibliography}{99}
\bibitem{kg6} See the website: www.slac.stanford.edu/spires/hep

\bibitem{kg6}  M. Morris and K. Thorne , American  J. Phys. 56, 39 (1988 )
\bibitem{kg10} M. Visser , Lorentzian Wormholes: From Einstien to
                   Hawking , AIP Press (1995)
\bibitem{kg10} Sharma. R and Maharaj. S, gr-qc/ 0702046 ( Mon.Not.Roy.Astron.Soc.375:1265-1268,2007 and references therein. )
 \bibitem{kg10} Usov.V , Phys.Rev.D 70, 067301 (2004)
\bibitem{kg10} Bronnikov K A et al, Class.Quan.Grav. 20, 3797
(2003); Tiwari R et al, Ind.J.Pure and Appl.Maths., 27, 907
(1996); Wang P et al,Class.Quan.Grav. 20, 3797 (2005); Sola J et
al, Mod.Phys.Lett.A, 21, 479(2006)
\end{thebibliography}
\end{document}